\documentclass[12pt]{article}

\usepackage{scicite}
\usepackage[utf8]{inputenc} % allow utf-8 input
\usepackage[T1]{fontenc}    % use 8-bit T1 fonts
\usepackage{hyperref}       % hyperlinks
\usepackage{url}            % simple URL typesetting
\usepackage{booktabs}       % professional-quality tables
\usepackage{amsfonts}       % blackboard math symbols
\usepackage{nicefrac}       % compact symbols for 1/2, etc.
\usepackage{microtype}
\usepackage{lipsum}
\usepackage{graphicx}
\usepackage{indentfirst}
\usepackage{amsmath}
\usepackage{multirow}
\usepackage{subfigure}
\usepackage{amsthm}
\usepackage{amssymb}
\usepackage{setspace}
\usepackage{authblk}
\usepackage{times}
\newenvironment{sciabstract}{%
\begin{quote} \bf}
{\end{quote}}

\title{Evolutionary games and spatial periodicity}

\author
{Te Wu$^{1}$, Feng Fu$^{2^*}$ \& Long Wang$^{3^*}$\\
\normalsize{$^{1}$Center for Complex Systems, Xidian University, Xi'an, China}\\
\normalsize{$^{2}$Department of Mathematics, Dartmouth College,
Hanover, New Hampshire, United States of America}\\
\normalsize{$^{3}$Center for Systems and Control, College of Engineering, Peking University, Beijing, China}\\
\normalsize{$^*$Correspondence and requests for materials should be addressed to F.F. (email: fufeng@gmail.com) and L.W. (email:
 longwang@pku.edu.cn).}
}

\date{}

\begin{document}
\maketitle
\baselineskip24pt

\begin{sciabstract}
We establish a theoretical framework to address evolutionary dynamics of spatial games under strong selection. As the selection intensity tends to infinity, strategy competition unfolds in the deterministic way of winners taking all. We rigorously prove that the evolutionary process soon or later either enters a cycle and from then on repeats the cycle periodically, or stabilizes at some state almost everywhere. This conclusion holds for any population graph and a large class of finite games. This framework suffices to reveal the underlying mathematical rationale for the kaleidoscopic cooperation of Nowak and May's pioneering work on spatial games: highly symmetric starting configuration causes a very long transient phase covering a large number of extremely beautiful spatial patterns. For all starting configurations, spatial patterns transit definitely over generations, so cooperators and defectors persist definitely. This framework can be extended to explore games including the snowdrift game, the public goods games (with or without loner, punishment), and repeated games on graphs. Aspiration dynamics can also be fully addressed when players deterministically switch strategy for unmet aspirations by virtue of our framework. Our results have potential implications for exploring the dynamics of a large variety of spatially extended systems in biology and physics.
\end{sciabstract}

\section*{Introduction}
How cooperative behaviour emerges and is maintained is an enduring problem in evolutionary biology\cite{maynardsmith(1973)nature, may(1981)nature, axelrod(1984)book, axelrod(1981)science, nowak(2006)science, szabo(2007)pr, hardin(1968)science, nowak(1992)nature2, nowak(1993)nature3}. The Prisoner's Dilemma game has become a standard metaphor to elucidate the problem\cite{nowak(1992)nature, huberman(1993)pnas, nowak(1994)pnas, nowak(1994)ijbc, mukherji(1996)nature, langer(2008)jtb, szabo(1998)pre, nowak(1993)ijbc, herz(1994)jtb}. In the standard Prisoner's Dilemma game, two players simultaneously decide to cooperate or to defect in one encounter. When both players cooperate, they each obtain the payoff $R$. When both players defect, they each obtain the payoff $P$. When one cooperator is confronted with a defector, the former reaps the payoff $S$ while the later accrues the payoff $T$. With $T>R>P>S$, one can easily deduce that defection is the optimal choice. Had both players cooperated, group benefit would be maximized. Strategy conflict between individual and collective interests leads to social dilemma.

Several mechanisms\cite{nowak(2006)science} have been proposed to explain the emergence and maintenance of cooperative behaviour. Of particular note is the spatial structure, which prescribes the interaction and strategy dispersal patterns within a population\cite{nowak(1992)nature, santos(2008)nature, ohtsuki(2006)nature, allen(2017)nature, szabo(2007)pr, szabo(2002)prl}. Evolutionary graph theory provides a mathematical tool for exploring evolutionary dynamics on structured populations. Vertices stand for individuals and edges determine who interact with whom. Effects of graph topology on the evolution of cooperation have been intensively studied\cite{nowak(1992)nature, hauert(2004)nature, ohtsuki(2006)nature, santos(2008)nature, allen(2017)nature, hauert(2021)pnas, suqi(2019)pnas, santos(2006)pnas, fufeng(2010)jtb, wute(2009)pre, ohtsuki(2007)prl, ohtsuki(2007)jtb, santos(2005)prl}. Ref\cite{nowak(1992)nature} pioneered this research line.

The model considered in the pioneering work\cite{nowak(1992)nature} is simple while generates profound results. Players are placed on the sites of a two-dimensional, $n \times n$ square lattice. Players each interact with eight immediate neighbors and himself according to the simplified Prisoner's Dilemma game with $T=b$, $R=1$, and $P=S=0$. When players each deterministically adopt the strategy with the highest payoff of the nine interacting partners, evolutionary kaleidoscopes, deterministic chaos and the stable coexistence of cooperators and defectors are observed for appropriate parameter ranges\cite{nowak(1992)nature, nowak(1993)ijbc, nowak(1994)ijbc, ohtsuki(2007)jtb}. This pioneering work\cite{nowak(1992)nature} has catalyzed a large body of literature on evolutionary games on graphs\cite{huberman(1993)pnas, nowak(1994)pnas, nowak(1994)ijbc, nowak(1993)ijbc, herz(1994)jtb, mukherji(1996)nature, hauert(2001)prsb, hauert(2002)ijbc, eshel(1998)aer, langer(2008)jtb, hauert(2004)nature, hauert(2002)science, sigmund(1992)nature, killingback(1998)jtb, killingback(1996)prsb, brauchli(1999)jtb, ohtsuki(2006)nature, santos(2008)nature, allen(2017)nature, santos(2006)pnas, suqi(2019)pnas, hauert(2021)pnas, fufeng(2010)jtb, ohtsuki(2007)jtb, santos(2005)prl}. One class is mainly concerned with evolutionary game dynamics under weak selection\cite{ohtsuki(2006)nature, allen(2017)nature, nowak(2004)nature, suqi(2019)pnas, ohtsuki(2007)prl, ohtsuki(2007)jtb}. As game has just a small effect on reproductive success, evolutionary dynamics can be approximated theoretically. Mathematical results are thus established for regular graphs\cite{ohtsuki(2006)nature, ohtsuki(2007)prl, ohtsuki(2007)jtb}, where individuals each have the same number of neighbors. For heterogenous graphs, evolutionary game dynamics have only been studied using computer simulations. Until very recently, strategy selection conditions was obtained through calculating the coalescent time of random walks, applicable to any population structure\cite{allen(2017)nature}. Weak selection is still a key assumption to guarantee the applicability, just as the authors have pointed out that the selection intensity must be small to the extent that selective differences are sufficiently small\cite{allen(2017)nature}. A second class deals with evolutionary game dynamics on graphs under strong selection\cite{huberman(1993)pnas, nowak(1994)pnas, nowak(1994)ijbc, nowak(1993)ijbc, herz(1994)jtb, mukherji(1996)nature, hauert(2001)prsb, hauert(2002)ijbc, eshel(1998)aer, langer(2008)jtb, hauert(2004)nature, hauert(2002)science, sigmund(1992)nature, killingback(1998)jtb, killingback(1996)prsb, santos(2008)nature, santos(2006)pnas, fufeng(2010)jtb, santos(2005)prl, wute(2009)pre}. Computer simulations are still the main technique\cite{wute(2009)pre, santos(2005)prl, santos(2006)pnas}. Pair approximation is also sometimes used to predict evolutionary trends\cite{fufeng(2010)jtb, hauert(2004)nature}. Very special cases are also directly analyzed\cite{eshel(1998)aer}.

Since its publication\cite{nowak(1992)nature}, this pioneering work has aroused great interest  and extensive literature has sought to probe effects of graph topology on evolutionary game dynamics under weak selection or strong selection. However, theoretical foundations remain unknown for such rich dynamics induced by deterministic updating rule\cite{nowak(1992)nature, huberman(1993)pnas, nowak(1994)pnas, nowak(1994)ijbc, nowak(1993)ijbc, herz(1994)jtb, mukherji(1996)nature, hauert(2001)prsb, hauert(2002)ijbc, eshel(1998)aer, killingback(1998)jtb, hauert(2004)nature, hauert(2002)science, sigmund(1992)nature}. As reported\cite{nowak(1992)nature}, striking spatial patterns generally emerge in evolutionary processes. Defectors and cooperators persist indefinitely in such patterns. Furthermore, the frequency of cooperators always oscillates in the same small range and seemingly independent of initial conditions for $1.8<b<2$. Approximation made on the fraction of cooperators not only agrees well with numerical simulations for large symmetric patterns but also works for irregular chaotical patterns which puzzled the authors themselves\cite{nowak(1992)nature}. Questions naturally arise whether there exists a universal constant governing Prisoners' Dilemma interactions on a lattice\cite{huberman(1993)pnas}, defectors and cooperators
would coexist in a chaotic way permanently\cite{nowak(1992)nature, sigmund(1992)nature}, and why fraction of cooperators depends on initial conditions for some interval of $b$ while independent of for other intervals?

\section*{Evolution exhibits periodicity almost everywhere}
We attempt to develop a theoretical framework to address evolutionary game dynamics on graphs under deterministic updating rule. We consider games on any evolutionary graph of finite vertices. In the game $M$, two types of strategies, $A$ and $B$, are feasible to players. The game is symmetric so that four parameters suffice to describe the payoff matrix. An $A$ player would get $a_{11}$ and $a_{12}$ if he interacts with an $A$ player and with a $B$ player, respectively. One $B$ player would get $a_{21}$ and $a_{22}$ if he interacts with an $A$ player and with a $B$ player, respectively. We represent the game as $M=(a_{11}, a_{12}, a_{21}, a_{22})$. Denote by $\Omega$ the parameter region of all possible payoff elements $a_{11}, a_{12}, a_{21}, a_{22}$. The graph consists of an interaction graph and a replacement graph. Each vertex is occupied by one individual. Individuals interact with neighbors according to the interaction graph. The replacement graph specifies how individuals update strategy. We focus on deterministic updating rule. After playing games with neighbors as defined by the interaction graph and with self (self-interaction included), each individual would deterministically adopt the strategy of one neighbor as defined by the replacement graph or himself, depending on who has achieved
the highest payoff.

\textbf{Theorem 1} For any evolutionary game $M$ defined on the region $\Omega$, the evolutionary process would, after a transient phase of finite generations, either be stabilized at one state, or enter periodic cycle almost everywhere on $\Omega$.

Applying \textbf{Theorem 1} to degree-regular graphs, we can arrive at \textbf{Corollary 1}.

\textbf{Corollary 1} For any evolutionary Prisoner’s Dilemma game with the single parameter $u \in (0, 1)$, where every individual in the interaction graph $G_I$ has $k$ neighbors (self interaction included),  the evolutionary process would either be stabilized at one state, or enter periodic cycle almost everywhere on $(0, 1)$, and the corresponding set of measure zero is $\mathcal{ZP}=\{\frac{i}{1+k}|i=1,2,...,k\}$.

For proofs of \textbf{Theorem 1} and \textbf{Corollary 1}, please see \textbf{Appendix A} and \textbf{Appendix B} in Supplementary Information.

\section*{Analysis of evolutionary dynamics for small population size}
For evolutionary Prisoner's Dilemma on a square lattice as defined in Supplementary Information, we can obtain the points of the measure-zero set as $\mathcal{ZP}=\{\frac{9}{8}, \frac{8}{7}, \frac{7}{6}, \frac{6}{5}, \frac{5}{4}, \frac{9}{7}, \frac{4}{3}, \frac{7}{5}, \frac{3}{2}, \frac{8}{5}, \frac{5}{3}, \frac{7}{4},\frac{9}{5}\}$. Except for these $13$ values of $b$ in the interval $(1,2)$, the evolutionary dynamics would be equilibrated at some periodical cycle or one state. Thus we can classify all the population states on the basis of their attractors. States converging to the same attractor (periodical cycle or one state) belong to the same class. Four periodical cycles, full cooperation and full defection constitute the attractors of the $4 \times 4$ latticed population and naturally divide all $65536$ states into six classes. Through analyzing the properties of attractors, the evolutionary dynamics can be fully revealed. In Fig. 1 we illustrate the states of one class and their evolutionary paths to the absorbed periodical cycle. Other classes are shown as Figs. S1 and S2.

\section*{Analysis of evolutionary processes for two typical intervals $1.75<b<1.8$ and $1.8<b<2$}
As population size increases, the states expand exponentially. It is thus impossible to classify graphically all the states. We can still find typical attractors for different intervals of $b$ and reveal their main properties. Let us now present the results for intervals $1.75<b<1.8$ and $1.8<b<2$, which has attracted great interest in the pioneering work\cite{nowak(1992)nature}. The authors claimed that the evolution ($1.75<b<1.8$) induces an irregular but static pattern, and the equilibrium frequency of cooperators depends upon the initial conditions (Fig. 1a in ref\cite{nowak(1992)nature}). For $1.8<b<2$, cooperators and defectors coexist indefinitely in a chaotically shifting balance, and the equilibrium frequency of cooperators is (almost) completely independent of the initial conditions (Fig. 1b in ref\cite{nowak(1992)nature}). However, evolutionary dynamics for both cases have no essential difference according to our \textbf{Corollary 1}. The evolution would eventually either repeat some periodical cycle or be stabilized, implying that the equilibrium frequency of cooperators is uniquely determined by the initial conditions.

Further comparison reveals three specific differences in the evolutionary processes for $b=1.751$ and $b=1.951$. First difference is the convergence rate. In the former case, after transient phases of several to several tens of generations, the
evolutionary processes can swiftly enter periodical cycles or stabilize for $n=20$ (Fig. 4a) or $n=30$ (Fig. 4c). Even when the population consists of as many as $400 \times 400$ vertices, periodical cycles also emerge in several tens to thousands of generations (Fig. 4e). For $1.8<b<2$, the periodical cycles can still be found in hundreds to thousands of generations for $20 \times 20$ (Fig. 4b). As the population size increases to $30 \times 30$, however, it generally takes several millions to tens of millions of generations to enter periodical cycle or stabilize (Fig. 4d). Further growth of the population requires unbearably long time to find the attractors (Fig. 4f). In another two specific simulations, for example, we still have not observed the periodical cycles in $100000000$ generations for $n=40$ and $70000000$ generations for $n=50$ (not shown here). It is thus impossible to claim if these evolutionary processes have already been inside periodical cycles or still in the transient phases. Even so, the population would eventually be (dynamically) equilibrated. As to what this (dynamical) equilibrium is like, we leave this annoyingly time consuming question to ever-increasingly powerful computers.

The first difference can be further elaborated by different ways that the evolutionary processes converge. Under selection pressure, defectors attempt to invade cooperators and sometimes successfully. For $b=1.751$, however, expansion of defectors' territories constrains their further expansion. Once defectors form clusters, defectors would have less cooperator neighbors to exploit. Peripheral defectors would soon be replaced by their neighboring cooperators, who are reciprocated by cooperators inside cooperator clusters. As a result, cooperators quickly cluster and thus press defectors into line fragments. Strategy switching just happens along these line fragments. So the evolutionary processes just visit very limited numbers of patterns of the population, explicating the rapid convergence of the evolutionary processes (Figs. 2a-d, Figs. 3a-d). $b=1.951$ further enhances the advantage of defectors over cooperators. Defectors still have chance to spread even having less cooperator neighbors compared with $b=1.751$. This hardens the survival of cooperators. Only by forming very special local clusters can cooperators defend defectors' invasion.
These clusters move, rotate, or transshape, somewhat like wave propagation, leading to their expansion or fragmentation. Owing to such a great variety of local invasion dynamics, the evolutionary processes would visit astronomical numbers of patterns of the population before entering periodical cycles (Figs. 2e-h, Figs. 3e-h). Based on this analysis, we may predict that reducing the number of local cooperator clusters can prevent them from colliding, and thus shorten the time required to reach equilibrium state. Our simulation results support this prediction. When the populations are initialized with just $5\%$ cooperators, the evolutionary processes in some simulation runs converge very quickly (Figs. 5e-h).

In the third place, the periodical cycles generally differ. For $1.75<b<1.8$, cooperators form clusters. Defectors parasite such clusters by surrounding them in narrow stripes or even line fragments, a phenomenon especially apparent for larger populations (Figs. 5a-d). Only individuals along these stripes or lines can switch to different strategies. Cooperators inside clusters remain static. On the whole, both the pattern and the number of cooperators change from generation to generation, showing the property of periodical oscillation. For $b=1.951$, the periodical cycles contain a very few local cooperator clusters. These clusters each transform periodically and no collision happens between them. The fraction of cooperators remains constant while their distributions are ever-changing following periodical modes (Figs. 2h,3h,5h).

Though the differences, the evolutionary processes would always enter periodical cycles or stabilize. The equilibrium frequency of cooperators thus can be precisely computed by averaging over the patterns in one full periodical cycle. So it is of great necessity to find the periodical cycle.
Ref\cite{nowak(1992)nature} made an approximation on the equilibrium frequency based on the data of $300$ (Fig. 2a in ref\cite{nowak(1992)nature}) or $2000$ (Fig. 2b in ref\cite{nowak(1992)nature}) generations for $1.8<b<2$, and claimed this approximation agrees well with numerical results. This claim is worthy of more effort to check. It is uncertain if such windows are still in transient phases or inside periodical cycles. Our numerical results have clearly demonstrated that the evolutionary processes generally converge to periodical cycles of very short length after surprisingly long transient phases for the population of size $30 \times 30$. The equilibrium frequency of cooperators is much lower than the average over any window in the transient phase.

\section*{Evolutionary kaleidoscope is just a flash in the pan and would eventually disappear}
 It should be noted that even very simple and highly symmetric initial configuration can induce quite long transient phase. Starting with a defector at the center of a $99 \times 99$ square-lattice world of cooperators with fixed boundary conditions, we have exactly replicated the patterns at generations of $30$, $217$, $219$ and $221$ as Figs. 3a-d in ref\cite{nowak(1992)nature}, respectively. The spatial games do generate a rich variety of beautiful patterns-an evolutionary kaleidoscope. But these patterns are just a flash in the pan. They would eventually disappear. The evolution would repeat a periodic cycle of length $4$. The evolutionary process enters this periodical cycle the first time at the $1016833950$\emph{th} generation. After that, $16$ local cooperator clusters, each consisting of six tightly connected cooperators, perpetuate themselves by rotating in the world of defectors (Figs. 6a-d). The estimate of asymptotic fraction of cooperators, $12ln2-8(\approx 0.318 )$, made on the basis of a very large symmetric pattern, does not suffice to approximate the true fraction of cooperators $(96/9801\approx 0.0098)$ when the evolution enters periodical cycle\cite{nowak(1992)nature, mukherji(1996)nature}.

\section*{Disucssion}
It is then straightforward to elaborate on the complexity and underlying regularity of the patterns\cite{nowak(1992)nature}. When two initial configurations lead the population to different periodic cycles, patterns in their evolutionary trajectories must have no overlap. Even if two configurations could lead the population eventually to the same periodical cycle, patterns in their transient phases may differ, especially when one transient phase covers much more generations than the other. Once the evolution enters periodical cycle, regularity appears. Different values of $b$ may induce different sets of specific objects observed locally while they do not qualitatively change the periodic behavior of evolution.

We have also explored the evolutionary dynamics for $b$ lying in other intervals. Representative patterns inside periodical cycles are presented in Fig. S3. For $1<b<9/8$, a very few number of defectors survive in the sea of cooperators. These defectors are in total isolation. Indeed, such polymorphic patterns are static (Fig. S3b). A defector in isolation adheres to defection as it always scores highest among its eight cooperator neighbors, who are anchored by their cooperator neighbors with payoff of $9$ ($>8b$). For $9/8<b<8/7$, patterns inside periodical cycle see many clusters of size $3 \times 3$. At the center of these clusters are defectors. Eight individuals surrounding each center switch between cooperation and defection synchronously. As $8/7<b<9/5(b\notin \mathcal{ZP})$, defector clusters of size $3\times 3$ fade and defectors form lines. The larger $b$, the more connected the lines and the fewer the defector clusters. Invariably is periodical strategy switching of individuals along the lines (or defector clusters). Defectors cannot expand into larger clusters. This further explains why evolutions quickly converge for $b<9/5$. For $9/5<b<2$, different scenario emerges. Both cooperators and defectors locally cluster. Two kinds of clusters nibble away at each other. In most cases after long transient phases, the evolutions reach dynamical equilibria in which selection drives local cooperator patterns to change periodically. So even the fraction of cooperators appears time-invariable, some of local patterns are ever-changing. When individuals just interact with eight nearest neighbors and the self-interaction is precluded, similar patterns (Fig. S4b, S4c, S4d, S4f) inside periodical cycles are observed while corresponding intervals change. The evolutions generally enter periodical cycles quickly except for $8/5<b<5/3$, in which cooperators and defectors chase each other very long generations seemingly irregularly (Fig. S4e). Similar interpretations just presented for $9/5<b<2$ (self-interaction considered) can be invoked for this 'interesting' region\cite{nowak(1992)nature}.

In \textbf{Theorem 1}, we do not make any assumption on the type of the finite games in question. For other forms of the prisoner's dilemma our conclusions still hold while the dimensionality of measure and the measure-zero set should be modified accordingly. Not limited to prisoner's dilemma, our conclusions also hold for snowdrift game, another simple game which has also attracted great attention in theoretical and experimental studies\cite{hauert(2004)nature, killingback(1996)prsb, killingback(1998)jtb, hauert(2001)prsb}. In the seminar work\cite{hauert(2004)nature}, the authors have also explored the evolutionary dynamics using the 'best-takes-over' rule, and concluded that it is impossible to derive clear trends for the evolutionary dynamics. In actuality, the evolutionary processes would repeat some periodical cycles or be stabilized. Once the cycles or stable states are found, the evolutionary dynamics can be theoretically addressed. It should be noted that when the update rule is of stochasticity rather than deterministic, the evolutionary dynamics can be fundamentally altered. The classes of the states of the Markov chain characterizing the evolutionary process may differ greatly. This explains why predictions by pair approximation are rather poor when the deterministic rule is approximated by a continuous sigmoid function\cite{hauert(2004)nature}.

Our conclusion can be further extended to multi-player multi-strategy games\cite{hardin(1968)science, hauert(2002)science, szabo(2007)pr, sigmund(2010)nature, szabo(2002)prl, santos(2008)nature}. The public goods game offers a good formula of decision making in an interaction with possibly conflicting interests\cite{szabo(2007)pr}. In a typical public goods game, a cooperator contributes an endowment of $c=1$ to the pool. A defector contributes nothing. All the contributions are multiplied by an enhancement factor $r$ and then equally distributed among all $g$ group members, regardless of their contributions. The constraint $1 < r < g$ leads to the social dilemma. Consider the evolutionary public goods game on a structured population of finite size\cite{szabo(2002)prl, santos(2008)nature, hauert(2002)science}. Individuals have the same number $(=z)$ of neighbors. All the individuals each organize a public goods game. Each player’s payoff is the sum over the payoffs of the games this player and his neighbors have organized. A defector’s payoff must be of the form $C_1 \cdot r-z$. A cooperator’s payoff must be of the form $C_2 \cdot r$. The coefficients $C_1$ and $C_2$ depend upon the local compositions while independent of $r$. Because the local compositions of defectors and cooperators are finite, defector’s and cooperator’s payoffs must be different as long as $r$ does not lie in the set $\{r|C_1 \cdot r-z=C_2 \cdot r\}$ which just contains a finite number of values. Following similar reasoning as in proving \textbf{Theorem 1}, we can conclude that the evolution would either enter periodic cycle or be stabilized for such spatial public goods games. Our conclusion still holds even for spatial public goods games on heterogenous populations, when individual's total contribution is proportional to degree and when individual's fixed contribution is equally divided among the groups in which they are engaged\cite{santos(2008)nature}. The periodicity or stabilization allows us to theoretically explore the effects of punishment\cite{sigmund(2010)nature} and voluntary participation\cite{szabo(2002)prl, hauert(2002)science} in spatial public goods games.

There is no requirement of the connectivity of the interaction graph and the replacement graph\cite{ohtsuki(2007)jtb, santos(2005)prl}. When the interaction graph completely overlaps with the replacement graph, our conclusion can be used to explore evolutionary dynamics on population graphs of high dimensions, viscous populations\cite{nowak(1994)pnas}, and heterogeneous populations as described by scale-free graphs\cite{santos(2005)prl, santos(2008)nature}. When the interaction graph does not overlap with the replacement graph\cite{ohtsuki(2007)jtb}, our conclusion is still applicable. This allows us to analytically delve into their respective roles in evolutionary dynamics of concern. Besides imitation-based strategy updating, our work can be extended to explore the aspiration dynamics\cite{herz(1994)jtb, zhoulei(2021)nc, chenxiaojie(2008)pre} analytically on the condition that strategy producing satisfactory payoff is maintained while strategy bringing payoff lower than aspiration switched.

Spatial reciprocity\cite{nowak(2006)science} is a mechanism for the evolution of cooperation based on the aggregation of cooperators. Most of previous studies confirmed this claim by simulations under strong selection\cite{santos(2005)prl, santos(2006)pnas}. Our work provides a theoretical foundation for convergence and our results confirm the stabilization of spatial reciprocity. Viewing the population as a whole, our theorem ensures the periodicity or stabilization of evolutionary games on graphs with deterministic updating rule. We may now turn to classifying the finite states of the the population depending on whether they drive the population to periodical cycles or stable states. With this classification, the evolutionary dynamics can be fully analyzed. In this sense, our results may provide a closed-form solution to the fundamentally undecided problem induced by the simplest conceivable two-player games\cite{sigmund(1992)nature}. For multiplayer games, there is considerably less theoretical research on evolutionary dynamics of structured populations, as more intricate local coupling discounts the effectiveness of the commonly used mathematical techniques such as pair approximation\cite{ohtsuki(2006)nature, szabo(2002)prl, szabo(2007)pr, fufeng(2010)jtb} and coalescent theory of random walk\cite{allen(2017)nature}. This discovery of spatial periodicity of evolutionary games opens up new possibilities to rigorously explore the evolutionary dynamics of spatial multi-player games\cite{hauert(2002)science, sigmund(2010)nature}. Moreover, this new evolutionary principle makes it possible to explore the dynamics of many spatially extended systems such as the species coexistence and diversity\cite{hassell(1991)nature, doebeli(1998)pnas}, the Red Queen effect\cite{hauert(2002)science}, and the collaboration in manufacturing systems\cite{zhengdazhong(1989)cta}, in which local interactions generally lead to large-scale patterns\cite{nakamaru(1997)jtb, hassell(1991)nature, hauert(2002)science, zhengdazhong(1989)cta, may(1976)nature}.

\section*{Acknowledgments}
Financial support from NSFC (62036002, 62273226) is gratefully acknowledged. Te Wu is also supported by the Fundamental Research Funds for Central Universities, Xidian University (JB210414).
\section*{Author Contributions} All authors conceived the study, performed the analysis, discussed the results and wrote the manuscript.
\section*{Data availability} Data and code are available upon request to Feng Fu and Long Wang.
\section*{Competing Interests} The authors declare that they have no competing financial interests.
\newpage

\begin{figure}[h!]
    \centering
    \includegraphics[scale=0.9]{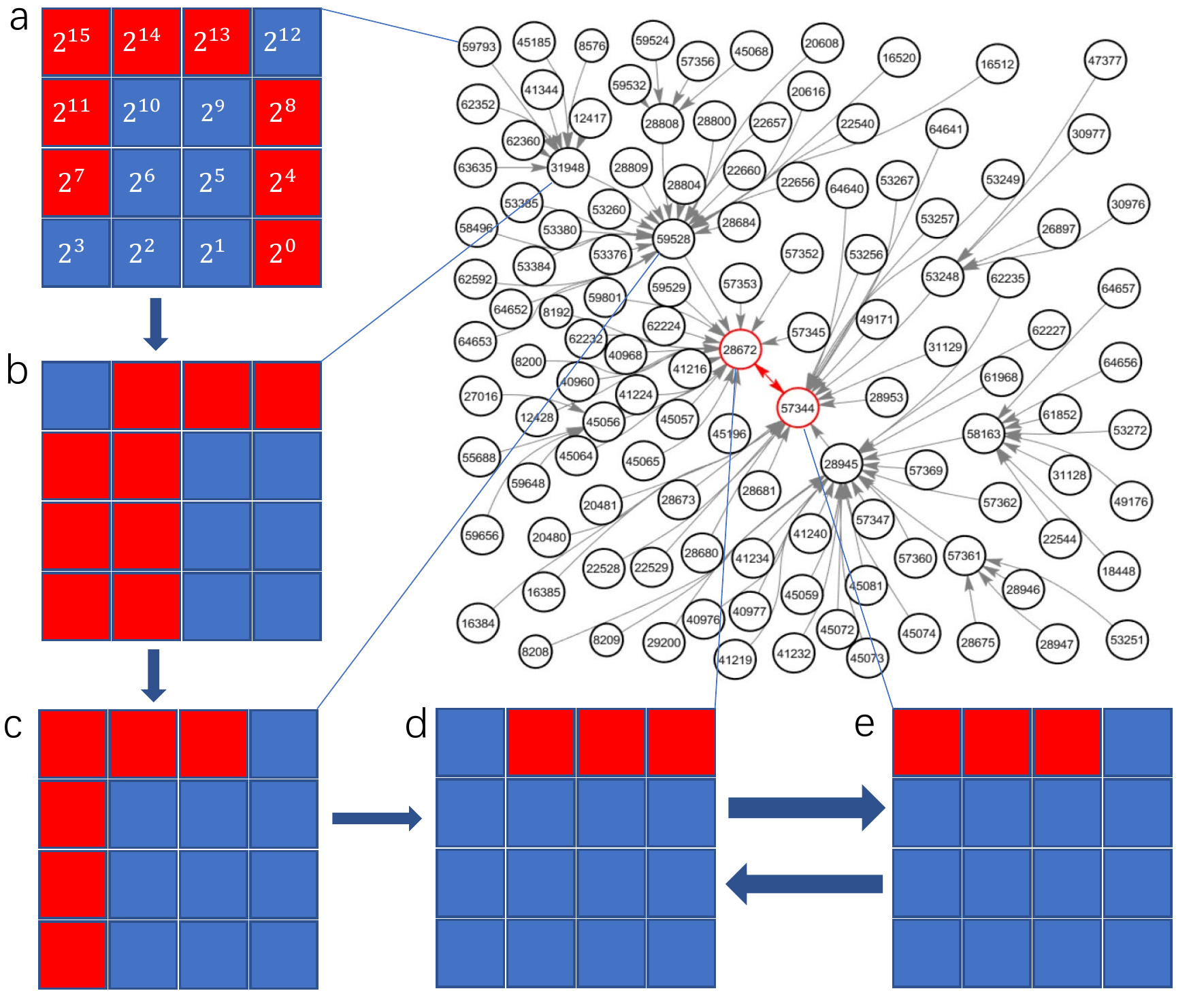}
    \caption{States of one class and their evolutionary paths to the absorbed periodical cycle. Starting with any state of this class, the evolutionary path is unique and the transition from generation to generation is completely deterministic. For this illustration, the evolution would be absorbed into the periodical cycle containing two states $28672$ and $57344$. Configurations a, b, c, d and e show a full evolutionary path starting from a. In these configurations, coloring Red denotes defection and Blue cooperation. Assign to Red $1$ and Blue $0$ in transforming  configuration to state (in decimal number), which is obtained by summing over each agent's strategy times its weight. Weights are as shown in a.}
\end{figure}

\newpage
\begin{figure}[h!]
    \centering
    \includegraphics[scale=1.5]{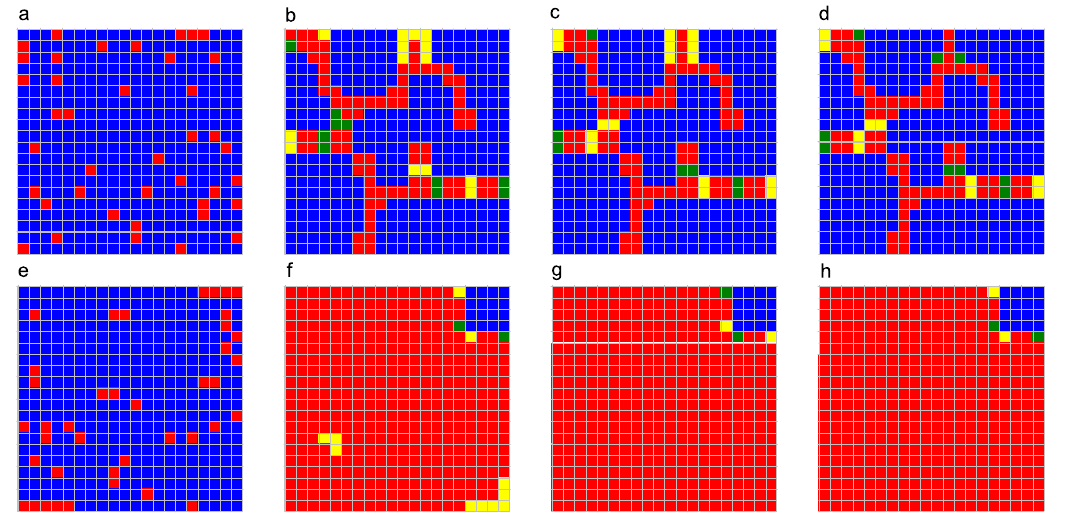}
    \caption{Under deterministic updating rule, the spatial Prisoner's Dilemma game would eventually either enter periodical cycle or stabilize after a transient phase, or short or long, depending on the population size, the starting pattern, and the magnitude of parameter $b$ representing the temptation to defect in the game. This figure shows two examples for $1.75<b<1.8$ and $1.8<b<2$ with the population structured by a $20\times 20$ square lattice with fixed boundary condition. The color coding is as follows. Blue represents a cooperator, initialized or following a cooperator. Red represents a defector, initialized or following a defector. Yellow a defector following a cooperator, and Olive a cooperator following a defector. For both ranges of $b$, the evolution can quickly enter periodical cycle. For $1.75<b<1.8$, the evolution starting from pattern a enters periodic cycle at generation $27$ and from then on repeat the cycle containing $12$ different patterns periodically. Pattern d is the last one of the first periodical cycle. The second periodical cycle starts from generation $39$, whose pattern would completely overlap with pattern b in terms of the distribution of cooperators and defectors. Pattern c is derived at generation $30$, also a typical pattern inside the periodical cycle. For $1.8<b<2$, the evolution starting from pattern e enters periodic cycle at generation $449$ (f) and from then on repeat the cycle containing $2$ different patterns periodically. So pattern h (generation $451$) completely overlaps with pattern f in terms of the distribution of cooperators and defectors. Pattern g is derived at generation $450$, also a typical pattern inside the periodical cycle.}
\end{figure}

\newpage
\begin{figure}[h!]
  \centering
  \includegraphics[scale=1.8]{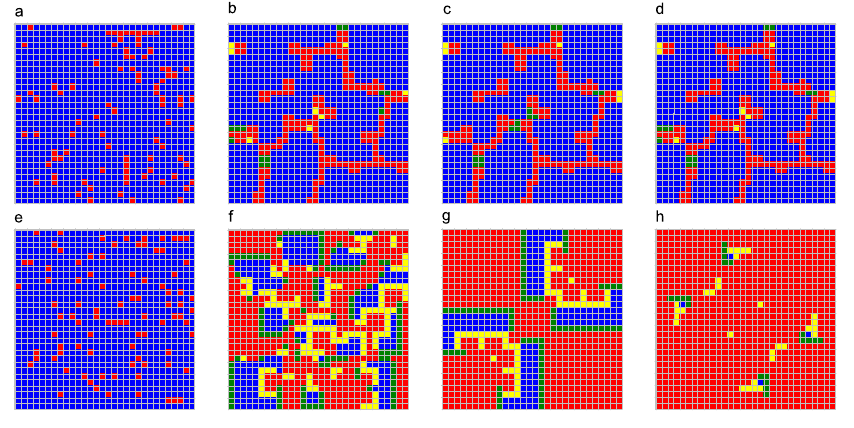}
  \caption{Evolutionary processes converge in quite different ways for $1.75<b<1.8$ and $1.8<b<2$ for the population as large as $30 \times 30$. The color coding is the same as in Figure 2. For $1.75<b<1.8$, cooperators form clusters quickly. Cooperators inside their clusters do not change strategy. Defectors persist in the form of line segments or narrow stripes interspersed between cooperator clusters. Only individuals along such interface between cooperator and defector are likely to switch strategies. Once defectors successfully assimilate their neighbors, defectors become victims of their success, as defectors have less cooperator neighbors to exploit. Marginal defectors along interface then become cooperators again. So the corresponding evolutionary process enters periodic cycle quickly (patterns a-d). Moving the temptation to the range $1.8<b<2$ aggravates the exploitation of defectors over cooperators. Especially when defectors have $\geq 5$ cooperator neighbors, they can always spread in such neighborhoods. Defectors can thus form clusters. Both cooperator clusters and defector clusters invade each other, leading to their expansion, collision and fragmentation. In general, only by forming very special local clusters can cooperators perpetually survive defectors' invasion (patterns e-h).}
\end{figure}

\newpage
\begin{figure}[h!]
\centering
\includegraphics[scale=0.6]{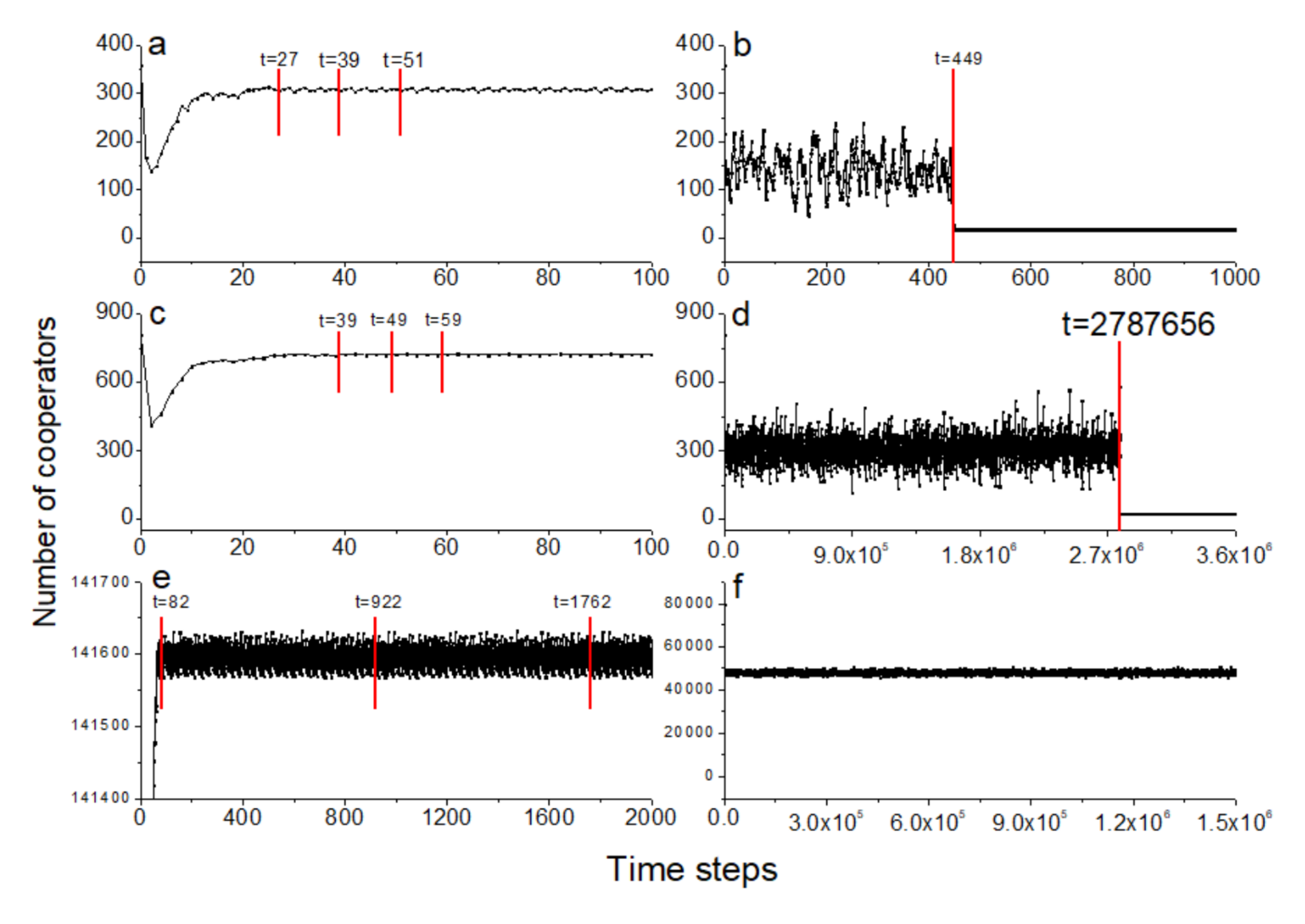}
\caption{Convergence rates present significantly different sensitivity to population size for $1.75<b<1.8$ and $1.8<b<2$. Evolution of spatial Prisoner's Dilemma can converge in several tens to thousands of generations for $1.75<b<1.8$ as the population size increases from $20\times 20$ to $30\times30$ to $400\times400$ (a,c,e). For $1.8<b<2$, the convergence rate is remarkably sensitive to the population size. For a $20\times 20$ population, the evolution can enter periodical cycle in hundreds of generations (b). As the population size grows to $30 \times 30$, it takes generally generations of magnitude $10^6$ to $10^7$ to enter periodical cycle (d). For a population as large as $400 \times 400$, the periodical cycle is still not found in $1.5\times 10^6$ generations. We thus cannot claim if these generations are already inside periodical cycle or still in transient phase. The most left vertical red line represents the generation at which the periodical cycle starts. Generations between two adjacent red lines constitute a full periodical cycle. In panel d, we record one value every $100$ generations for such drastic oscillation.}
\end{figure}

\newpage
\begin{figure}[!h]
  \centering
  \includegraphics[scale=1.8]{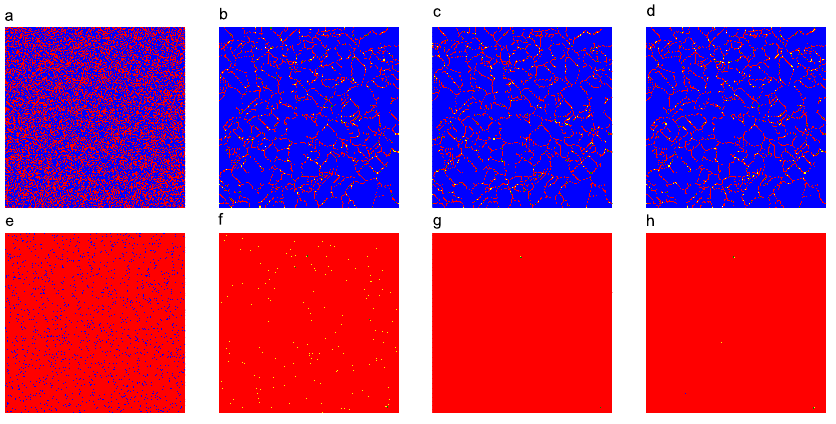}
  \caption{The spatial evolutionary game still quickly enters periodical cycle for $1.75<b<1.8$ for population as large as $400\times 400$. The color coding is the same as in Figure 2. One typical example shows that the evolution, starting with $60\%$ cooperators randomly distributed in the population (pattern a), enters periodical cycle at generation $82$ (pattern b). Visiting $840$ different patterns similar to pattern c, the evolution starts to repeat the periodical cycle at generation $922$ (pattern d). For $1.8<b<2$, cooperator clusters and defector clusters locally collide and cascade. The evolution generally transits extremely numerous patterns before entering periodical cycle. However, reducing the initial fraction of cooperators lowers the chance of collision of cooperator and defector clusters in the evolutionary process, and the evolution in some cases can enter periodical cycle quickly. An example shows that the evolution starting with a random distribution of $5\%$ cooperators (pattern e), can enter periodical cycle at the $4$th generation (pattern h). Patterns f and g confirm no collision in the transient phase. Most of cooperators are eliminated by defectors. Only three separated cooperator clusters survive the evolutionary race by rotating periodically.}
\end{figure}

\newpage
\begin{figure}[!h]
 \centering
\includegraphics[scale=2]{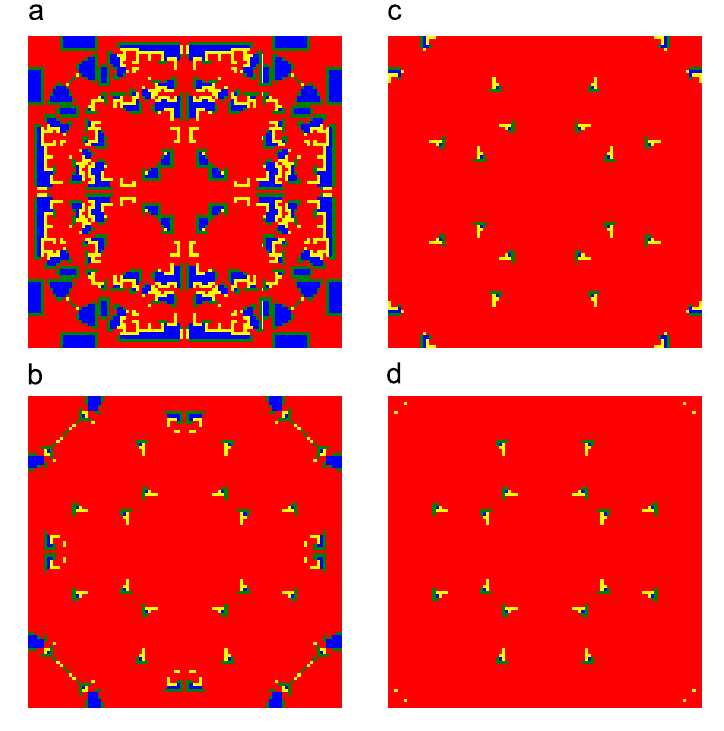}
\caption{Very simple initial configuration can cause extremely long transient phase for $1.8<b<2$. This figure shows a typical evolutionary process starting with a defector at the center of a $99 \times 99$ square-lattice world of cooperators with fixed boundary conditions. The color coding is the same as in Figure 2. The transient phase contains a very large number of different patterns, as many as $1016833950$, the initial pattern included. These patterns are all a flash in the pan. They would eventually disappear. Specific patterns at generation $1016833910, 1016833930, 1016833940$ are shown as a, b, c and d. One can even by hand check that starting from pattern d, the evolution would repeat a periodical cycle of length $4$. From then on the fraction of cooperators is $96/9801\approx0.0098$ and does not change as the evolution advances, while $12$ local cooperator clusters are ever-changing periodically.
}
\end{figure}

\newpage
\textbf{
Supplementary Information (SI)}

% Place the author information here.  Please hand-code the contact
% information and notecalls; do *not* use \footnote commands.  Let the
% author contact information appear immediately below the author names
% as shown.  We would also prefer that you don't change the type-size
% settings shown here.

\date{}
\maketitle
Section 1.1 presents evolutionary graphs used to structure populations. Section 1.2 defines how to compute payoffs of individuals located on vertices of interaction graph. Section 1.3 presents how to update strategies.
Section 2 defines evolutionary games on graphs. Our main result is presented as \textbf{Theorem 1}. Applying \textbf{Theorem 1} to degree regular graphs we obtain \textbf{Corollary 1}. Section 3 illustrates how we apply \textbf{Theorem 1} to classic settings where both interaction graph and replacement graph are characterized by the square lattice with fixed boundary conditions. Proofs of \textbf{Theorem 1} and \textbf{Corollary 1} are presented as \textbf{Appendix A} and \textbf{Appendix B}, respectively.

\section{Three elements for evolutionary games}
Evolutionary graph (interaction graph and replacement graph), payoffs and strategy updating constitute three elements of evolutionary games on graphs.
\subsection{Evolutionary graphs} An evolutionary graph $G=(V, E_I, E_R)$ consists of an interaction graph $G_I=(V, E_I)$ and a replacement graph $G_R=(V, E_R)$. The set $V$ is finite and consists of $N$ vertices. Label each vertex with a number and we can write the set $V$ as $\{1,2,3,...,N\}$. By undirected we mean that the edges are denoted by unordered pairs of nodes. For $u, v \in V$ and $u \neq v$, the pair $(u,v)$ denotes an unordered edge and node $u(v)$ is called a neighbor of node $v(u)$. Denote by $\mathcal{G}$ the set of all $\frac{N(N-1)}{2}$ unordered edges. Both $E_I$ and $E_R$ are subsets of $\mathcal{G}$. Define the degree $k_v^V$ of node $v$ in graph $V$ as the number of neighbors of node $v$. Interactions of individuals happen along edges in the graph $G_I$. By \textbf{self-interaction included} we mean that every individual also interacts with himself one time. Strategy imitation takes place along the edges in the graph $G_R$. The graph $G_I$ thus determines the payoffs of individuals and the graph $G_R$ underlies the structure how strategies spread.
\subsection{Payoff of individuals} Each vertex is occupied by an individual. Interaction is pairwise. Symmetric two-player game is involved to describe the interaction. Two strategies A and B are feasible to each player. Player $A$ will get the payoff $a_{11}$ when he interacts with an A player, and $a_{12}$ when with a B player.
Player B will get the payoff $a_{21}$ when he interacts with an A player, and $a_{22}$ when with a B player. So the payoff matrix $M$ is defined as follows: \begin{table}[!htbp]\large
\centering
\begin{tabular}{ccccc}
  & & $A$ & $B$ & \\
$A$ & \multirow{2}*{$\bigg($} & $a_{11}$ & $a_{12}$ & \multirow{2}*{$\bigg)$} \\
$B$ & & $a_{21}$ & $a_{22}$ & \\
\end{tabular}
\end{table}

In an evolutionary game, each individual interacts with his neighbors in the interaction graph and accumulates payoff. Denote by $pay(x,y)$ the payoff of individual $x$ when $x$ interacts with $y$. Then in the interaction graph the payoff of individual $x$ would be $p(x)=\sum_{(x,y)\in E_I}pay(x,y)$.
\subsection{Deterministic strategy updating} Strategy updating happens along the edges in the replacement graph. In the deterministic strategy updating, each individual would adopt the strategy of his neighbor who has obtained the highest payoff. Synchronous updating means that all the individuals update their strategies simultaneously.

\section{Evolutionary games} Given the game and the interaction graph $G_I$ and the replacement graph $G_R$, let us now describe how the system evolves according to the deterministic updating rule. The evolutionary process consists of initialization and iteration. In the initialization, a fraction of individuals are assigned to be A players and the rest of B players. With the initialization, the evolutionary process repeats two steps which we call a generation. In step $1$, all individuals each accumulate payoff by interacting with their neighbors in $G_I$. To make the payoff of every individual well-defined, we assume that every individual has at least one neighbor in the interaction graph. If self-interaction is included, this assumption is dispensable. In step $2$, individuals synchronously update their strategies deterministically. Specifically, each individual will imitate the strategy of his neighbor (in $G_R$) or himself with the highest payoff. For such evolutionary game on graph, one question of great concern is if there exists a general principle governing the evolutionary process?

For realistic reasons, we can reasonably assume $a_{11}\in I_{11}$, $a_{12}\in I_{12}$, $a_{21}\in I_{21}$, $a_{22}\in I_{22}$. Here $I_{11}$, $I_{12}$,
$I_{21}$, $I_{22}$ are intervals of finite length respectively and they are independent of each other. Then $\Omega=I_{11}\times I_{12}\times I_{21} \times I_{22}$ constitutes the parameter region of all possible payoff elements. To present our main results in a more transparent way, the parameter region of the game in question is measured by the Lebesgue integral\cite{zorich(2004)book} of the constant function
$\textbf{1}(\mu_1, \mu_2, \mu_3, \mu_4)$ on the region. Given the parameter region $\Omega$ we have its measure as $\mu(\Omega)=\int_\Omega \textbf{1}(\mu_1, \mu_2, \mu_3, \mu_4)\cdot d\mu_1 d\mu_2 d\mu_3 d\mu_4$. We say a set $I$ has measure zero\cite{zorich(2004)book} if $\mu(I)=0$. In line with the definition in measure theory\cite{zorich(2004)book}, we say that a property holds almost everywhere on $\Omega$ if this property holds at all points of the set $\Omega$ except possibly the points of a set of measure zero. Next we present our main results as Theorem 1 and Corollary 1.
\section*{Theorem 1} For any evolutionary game $M$ defined on the region $\Omega$, the evolutionary process would, after a transient phase of finite generations, either be stabilized at one state, or enter periodic cycle almost everywhere on $\Omega$.

 \textbf{Remark}: \textbf{Theorem 1} reveals the general principle that governs the evolutionary processes for \textbf{evolutionary games on graphs}, and also provides a theoretical foundation for the convergence of the evolutionary processes. Using this theorem, we can explore the evolution of social behavior on any population structure such as the evolution of cooperation in both Prisoner's Dilemma game and the Snowdrift game, which have attracted most attentions in theoretical and experimental studies. Previous theoretical studies are mainly based on approximations, and sometimes they just make a very poor prediction of evolutionary trends and even fails. Without convergence proof, results based on computer simulations are not always persuasive. In the \textbf{single parameter Prisoner's Dilemma game}, just a parameter $u$ is used to define the payoff matrix with $a_{11}=1, a_{12}=0, a_{21}=1+u, a_{22}=u$ and $u\in (0,1)$. When every individual in the interaction graph has $k$ neighbors, our main conclusion still holds and the measure-zero set can be explicitly obtained. Now let us present this result as \textbf{Corollary 1}.

 \section*{Corollary 1} For any evolutionary Prisoner’s Dilemma game with the single parameter $u \in (0, 1)$, where every individual in the interaction graph $G_I$ has $k$ neighbors (self interaction included),  the evolutionary process would either be stabilized at one state, or enter periodic cycle almost everywhere on $(0, 1)$, and the corresponding set of measure zero is $\mathcal{ZP}=\{\frac{i}{1+k}|i=1,2,...,k\}$.

   \textbf{Remark}: Our conclusion still holds when self-interaction is precluded while the measure-zero set should be modified as $\mathcal{ZP}=\{\frac{i}{k}|i=1,2,...,k-1\}$.
 \section{Evolutionary Prisoner's Dilemmas on square lattices}
 When both interaction graph and replacement graph are characterized by the square lattice with fixed boundary conditions, we can obtain the measure-zero set more explicitly. In the square lattice (Moore neighborhood) with fixed boundary conditions, vertices located inside the boundary each have eight neighbours. Vertices sitting on the boundary have five or three neighbors, depending on their specific locations. In line with ref\cite{nowak(1992)nature}, we consider the simplified Prisoner's Dilemma games with $a_{11}=1, a_{12}=0, a_{21}=b, a_{22}=0$ and $b\in (1,2)$.  For an evolutionary game on the square lattice with fixed boundary conditions, the evolutionary process would either be stabilized at some state or cycle on some period except a measure-zero set of $b$. Now let us show how this measure-zero set is like. We can easily write down the set of possible payoff for a cooperator as $\mathcal{P}^C=\{1,2,3,4,5,6,7,8,9\}$ and that for a defector as $\mathcal{P}^D=\{0,b,2b,3b,4b,5b,6b,7b,8b\}$. To make the intersection of the sets $\mathcal{P}^C$ and $\mathcal{P}^D$ non-empty, the parameter must take one of those values $\frac{3}{2}, \frac{4}{3}, \frac{5}{3}, \frac{5}{4}, \frac{7}{4}, \frac{6}{5}, \frac{7}{5}, \frac{8}{5}, \frac{9}{5}, \frac{7}{6}, \frac{8}{7}, \frac{9}{7},\frac{9}{8}$. Arranging these values from small to large, we get the set $\mathcal{ZP}$ as $\{\frac{9}{8}, \frac{8}{7}, \frac{7}{6}, \frac{6}{5}, \frac{5}{4}, \frac{9}{7}, \frac{4}{3}, \frac{7}{5}, \frac{3}{2}, \frac{8}{5}, \frac{5}{3}, \frac{7}{4},\frac{9}{5}\}$, we have  $\mu(\mathcal{ZP})=\int_{\mathcal{ZP}}\textbf{1}(b)d b=0$. For any value of $b\in (1,2)$ but not belonging to the set $\mathcal{ZP}$,  $\mathcal{P}^C \bigcap \mathcal{P}^D=\emptyset$, implying that cooperators and defectors always obtain different payoffs. For the deterministic strategy updating rule, that each individual adopts cooperation or defection is a certain event. Viewing the population as a whole, the population state would transit deterministically from one to another. The finiteness of possible population states leads to either periodicity or stabilization of the evolution.

\section*{Appendix A: Proof of Theorem 1}
 Suppose the maximum degree of vertices in $G_I$ is $K$. Each vertex is occupied either by an $A$ player or a $B$ player. Denote by $s^i(t)$ the strategy that individual $i$ adopts at generation $t$. Then the population state at generation $t$ can be represented as  $s(t)=(s^1(t),s^2(t),...,s^N(t)) (t=0,1,2,...)$.
 Let $\mathcal{P}^A=\{a_{11}+a_{11}\cdot k+a_{12}\cdot (m-k)|k=0,1,2,...,m;m=0,1,2,...,K\}$ and $\mathcal{P}^B=\{a_{22}+a_{21}\cdot k+a_{22}\cdot (m-k)|k=0,1,2,...,m;m=0,1,2,...,K\}$. It is obvious that
  the set of all possible payoffs for an $A$ player in the interaction graph must be a subset of $\mathcal{P}^A$, and the set of all possible payoffs for a $B$ player in the interaction graph must be a subset of $\mathcal{P}^B$. Let $\mathcal{ZP}=\{(a_{11},a_{12}, a_{21}, a_{22})|\mathcal{P}^A\cap\mathcal{P}^B\neq\emptyset\}$ and
  $\mathcal{NP}=\Omega\backslash\ \mathcal{ZP}$. As both $\mathcal{P}^A$ and $\mathcal{P}^B$ are finite sets, one can easily check that the set $\mathcal{ZP}$ has measure zero. In fact, for any $(a_{11},a_{12}, a_{21}, a_{22})\in \mathcal{ZP}$, there must exist integers $k_1, m_1, k_2, m_2$
  such that $a_{11}+a_{11}\cdot k_1+a_{12}\cdot (m_1-k_1)=a_{22}+a_{21}\cdot k_2+a_{22}(m_2-k_2)$. Rewriting this equation,
  we get a subset $\Omega'$ of the parameter region $\Omega$ as $\Omega'=\{(a_{11}, a_{12}, a_{21}, a_{22})|(k_1+1)a_{11}+(m_1-k_1)a_{12}-k_2 a_{21}-(1+m_2-k_2)a_{22}=0\}$.
  We thus have $\mu(\Omega')=\int_{\Omega'}\textbf{1}(\mu_1, \mu_2, \mu_3, \mu_4)\cdot d\mu_1 d\mu_2 d\mu_3 d\mu_4=0$, since one entry is always uniquely determined by the other three entries of any tuple lying in $\Omega'$. According to the additivity of measure, we have
 $\mu(\mathcal{ZP})=0$ because the set $\mathcal{ZP}$ consists of a finite number of such $\Omega'$s.
   Given any tuple $(a_{11},a_{12}, a_{21}, a_{22})\in \mathcal{NP}$, we have $\mathcal{P}^A\cap\mathcal{P}^B=\emptyset$. This means that any two individuals sticking to different strategies in the interaction graph obtain different payoffs. As a result, any two individuals of different strategies in the replacement graph have different payoffs. So given any player in the replacement graph, either some $A$ player(s) obtain the highest payoff or some $B$ player(s) garner the highest payoff among his neighbors.
But these two cases are unlikely to happen at the same time. In other words, which strategy one player would adopt in next generation is a deterministic event. So the state transition of the evolutionary graph is also a deterministic event. Suppose the evolution starts with state $s(0)$. Denote by $s(1), s(2), s(3),..., s(2^N)$ the states at generation $1,2,3,...,2^N$. As the evolutionary graph consists of $N$ nodes, the graph admits at most $2^N$ different states. Using Dirichlet's Pigeonhole Principle\cite{peter(1987)book}, we can conclude that there must be different generations $l^{'}(0\leq l^{'}\leq 2^N-1)$ and $h^{'}(1\leq h^{'}\leq 2^N)$ such that $s(l^{'})=s(h^{'})$. Let $l$ and $h$ be the minimals of all such $l^{'}$s and of all such $h^{'}$s, respectively. If $h=l+1$, the evolution would be stabilized at the state $s(l)$. If $h\geq l+2$, the evolution would enter periodical cycle from the $l$\emph{th} generation and from then on cycle as $s(l), s(l+1), ...,s(h-1), s(l),s(l+1),...,s(h-1),...$. The states $s(l), s(l+1),...,s(h-1)$ constitutes a full periodical cycle and $h-l$ is the minimal period. The proof is thus completed.

\section*{Appendix B: Proof of Corollary 1}
 The set of possible payoffs for a cooperator is $\mathcal{P}^C=\{1,2,...,1+k\}$. The set of possible payoffs for
 a defector is $\mathcal{P}^D=\{(k+1)u, (1+u)+ku, 2(1+u)+(k-1)u,...,k(1+u)+u\}$. For $u\in (0,1)$, to ensure that the interaction of the sets $\mathcal{P}^C$ and $\mathcal{P}^D$ would not be
empty, $u$ must be one of these values $\frac{1}{k+1}, \frac{2}{k+2}, ...,\frac{k}{k+1}$, which can be
collected as the set $\mathcal{ZP}=\{\frac{i}{k+1}|i=1,2,...,k\}$. It is obvious that the set $\mathcal{ZP}$ is comprised
of a finite number of points and thus we have $\mu(\mathcal{ZP})=\int_{\mathcal{ZP}}\textbf{1}(\mu)d\mu=0$.
Following similar reasoning as in proving \textbf{Theorem 1}, we can conclude that the evolutionary process would either enter a periodical cycle or be stabilized almost everywhere on $(0,1)$ as the set $\mathcal{ZP}$ is still of measure zero.

%\begin{figure}
% \centering
%\includegraphics[scale=0.8]{illustration22.pdf}
%\caption{To present the evolutionary path in a transparent way, we map the population state (configuration) to a decimal number. Each vertices is associated with a weight as shown in a. Red denotes defection. Blue denotes cooperation. Assign to cooperation $1$ and defection $0$. The decimal number of a given configuration is obtained by summing over each vertex' number times its weight. We can thus easily get states of configurations a, b and c as $51231$, $15$ and $0$ (numbers inside circles). Under deterministic updating rule, the evolutionary process starting with a will transit to b and then to c and stay in c forever. The simplified Prisoner's Dilemma game is used to characterize the pairwise interaction. Self-interaction is also included. Parameter $b$ lies in the interval $(1.75,1.80)$.}
%\end{figure}

\newpage
\begin{figure}
 \centering
\includegraphics[scale=0.62]{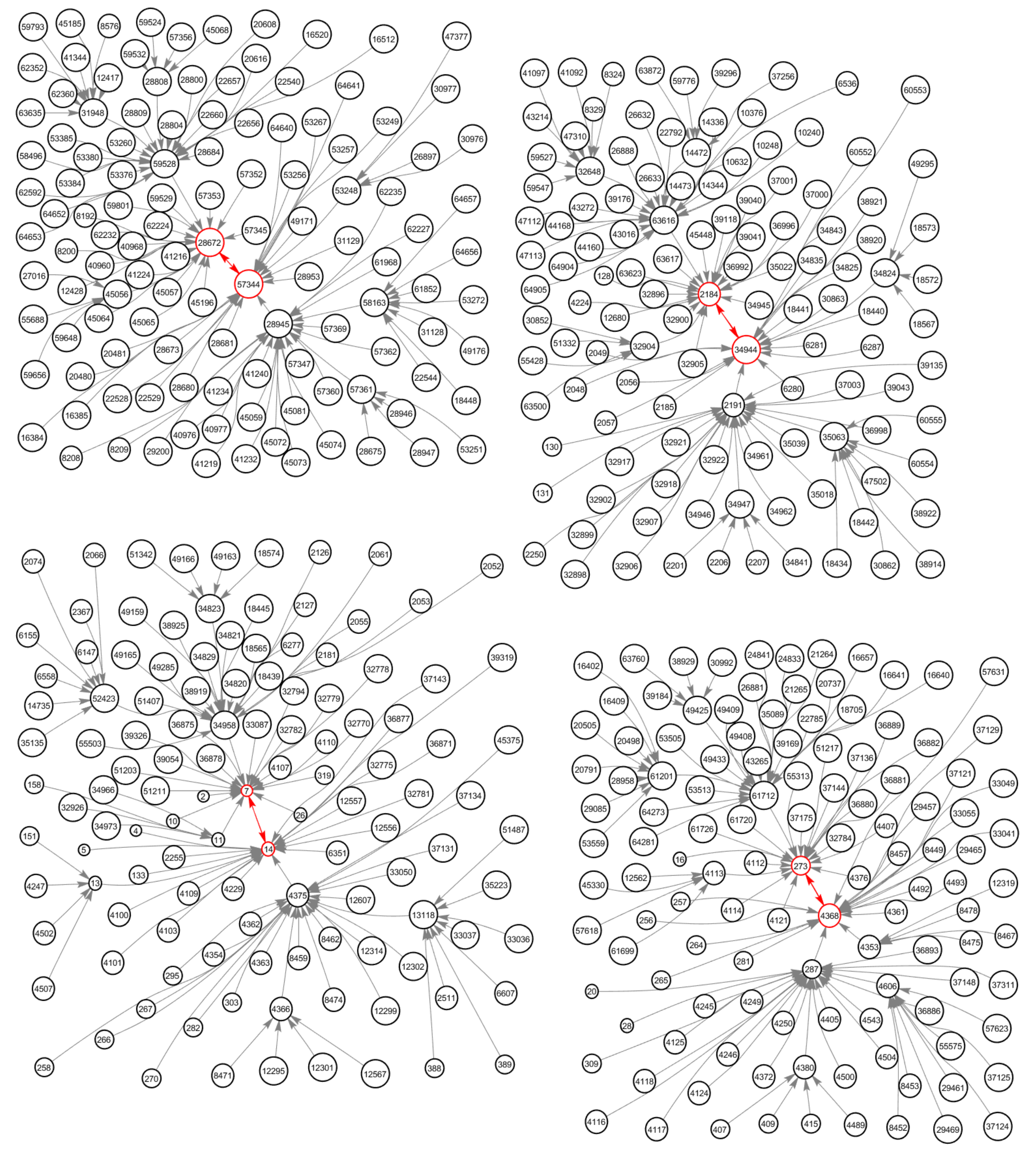}
%\caption
\begin{quote}
\textbf{Figure S1.} Four classes of states converging to periodical cycles. Each absorbing periodical cycle is colored Red and associated states would be absorbed to this cycle. Each population configuration is transformed into one state following the same way as in Figure 1.
\end{quote}

\end{figure}

\newpage
\begin{figure}[]
	\centering
	\vspace{-0.35cm}
	\subfigtopskip=2pt
	\subfigbottomskip=2pt
	\subfigcapskip=-5pt
	\subfigure{
		\label{level.sub.1}
		\includegraphics[width=0.48\linewidth]{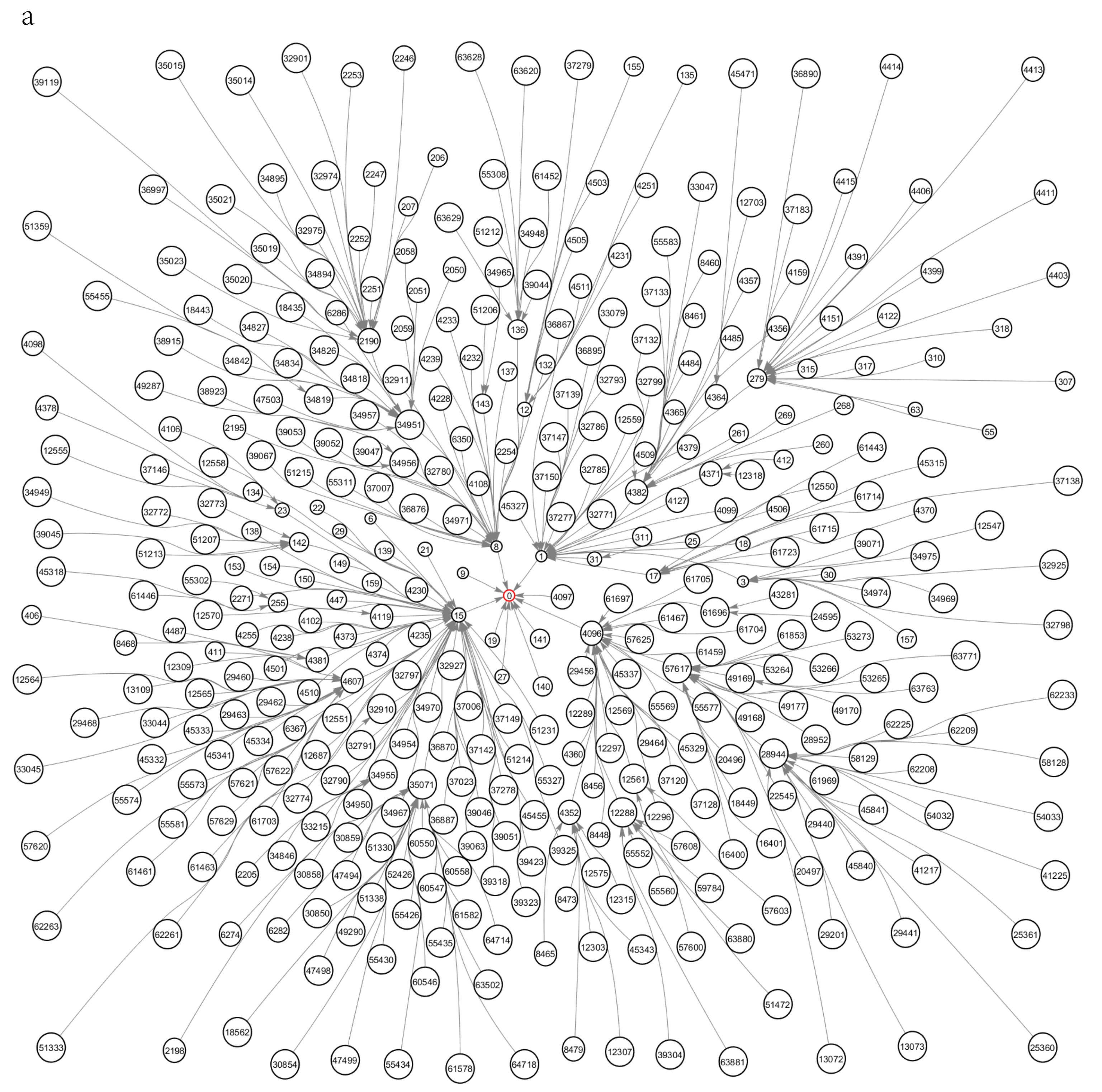}}
	\quad
	\subfigure{
		\label{level.sub.2}
		\includegraphics[width=0.4\linewidth]{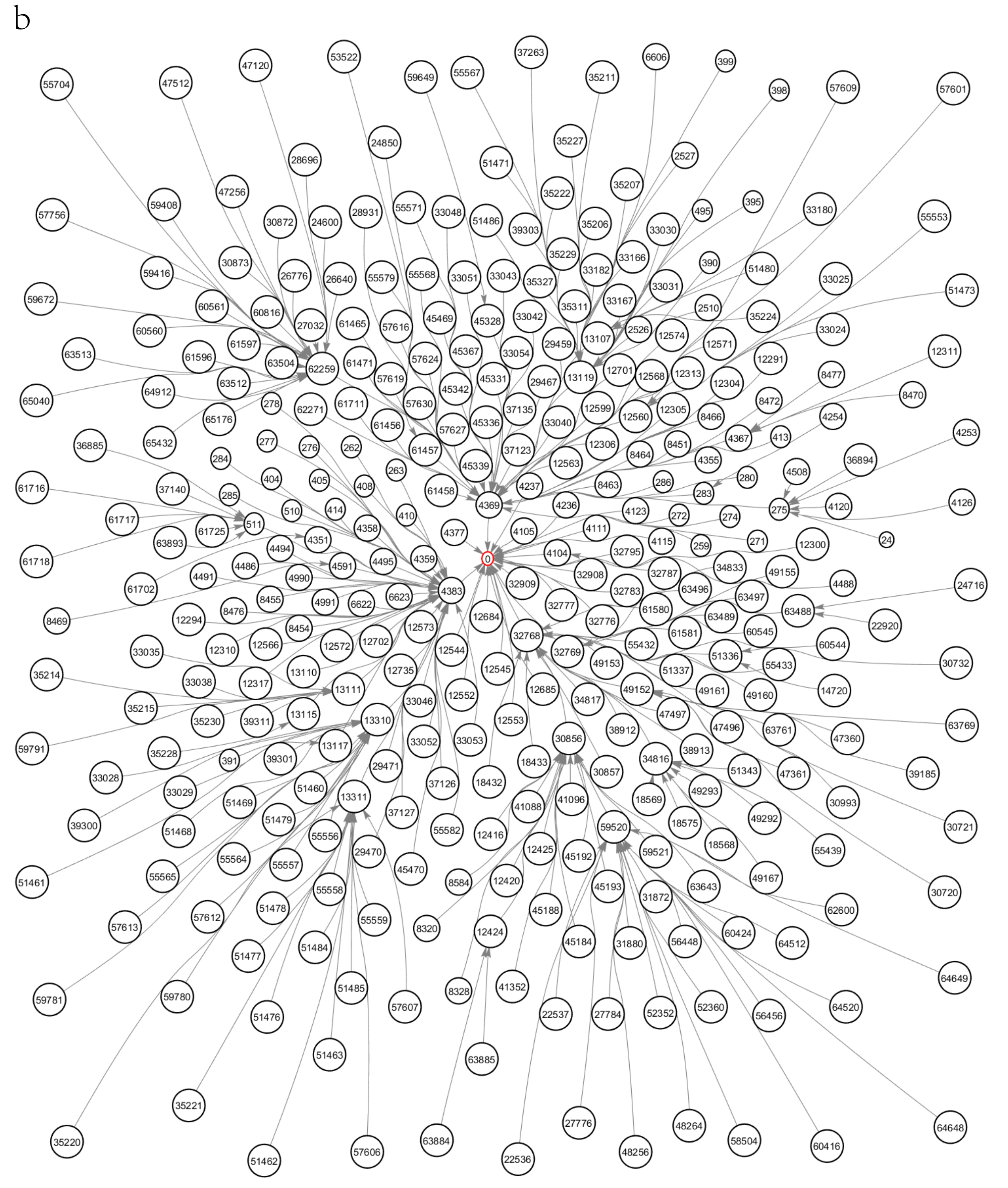}}
	\subfigure{
		\label{level.sub.3}
		\includegraphics[width=0.45\linewidth]{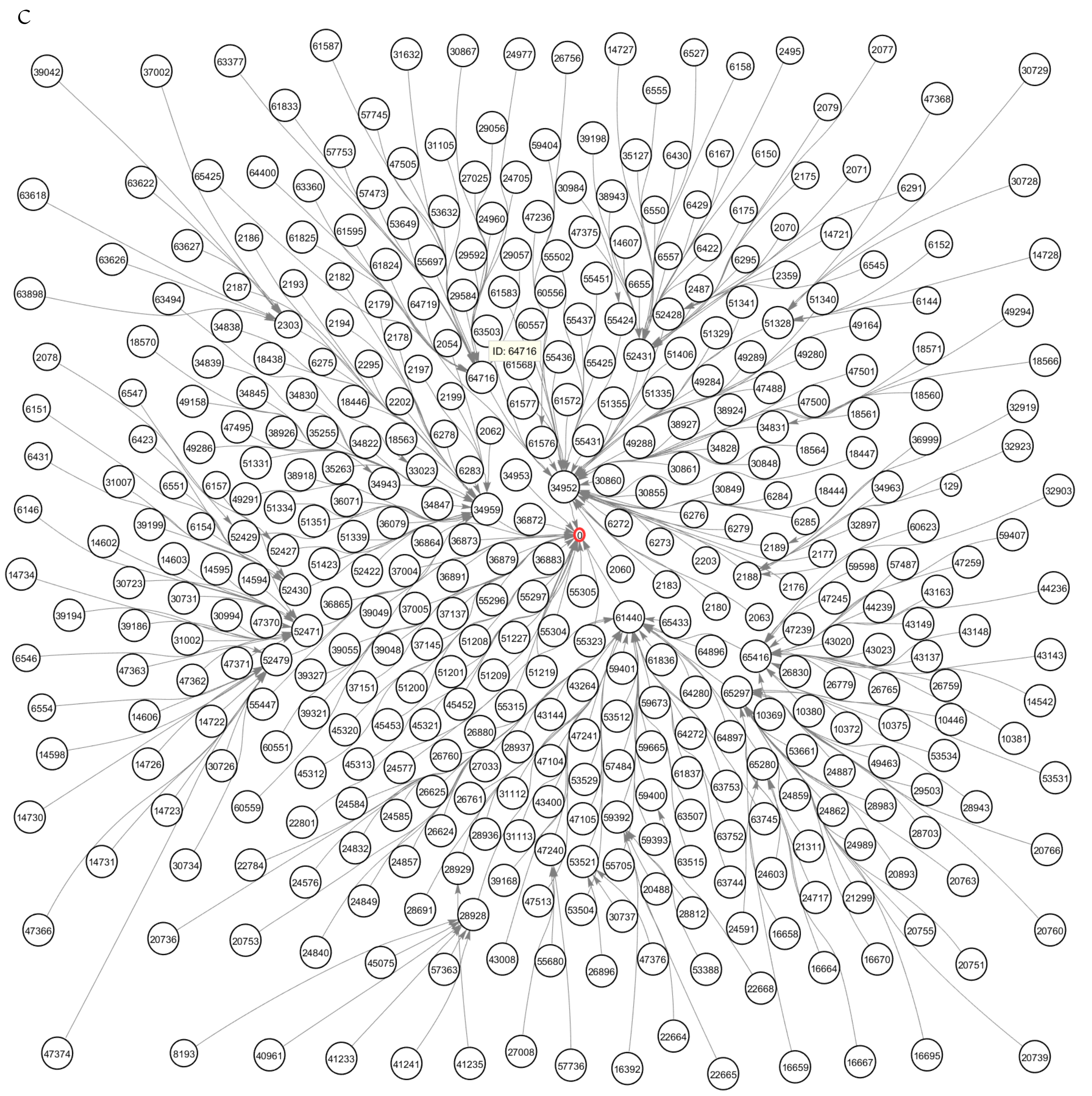}}
	\quad
	\subfigure{
		\label{level.sub.4}
		\includegraphics[width=0.45\linewidth]{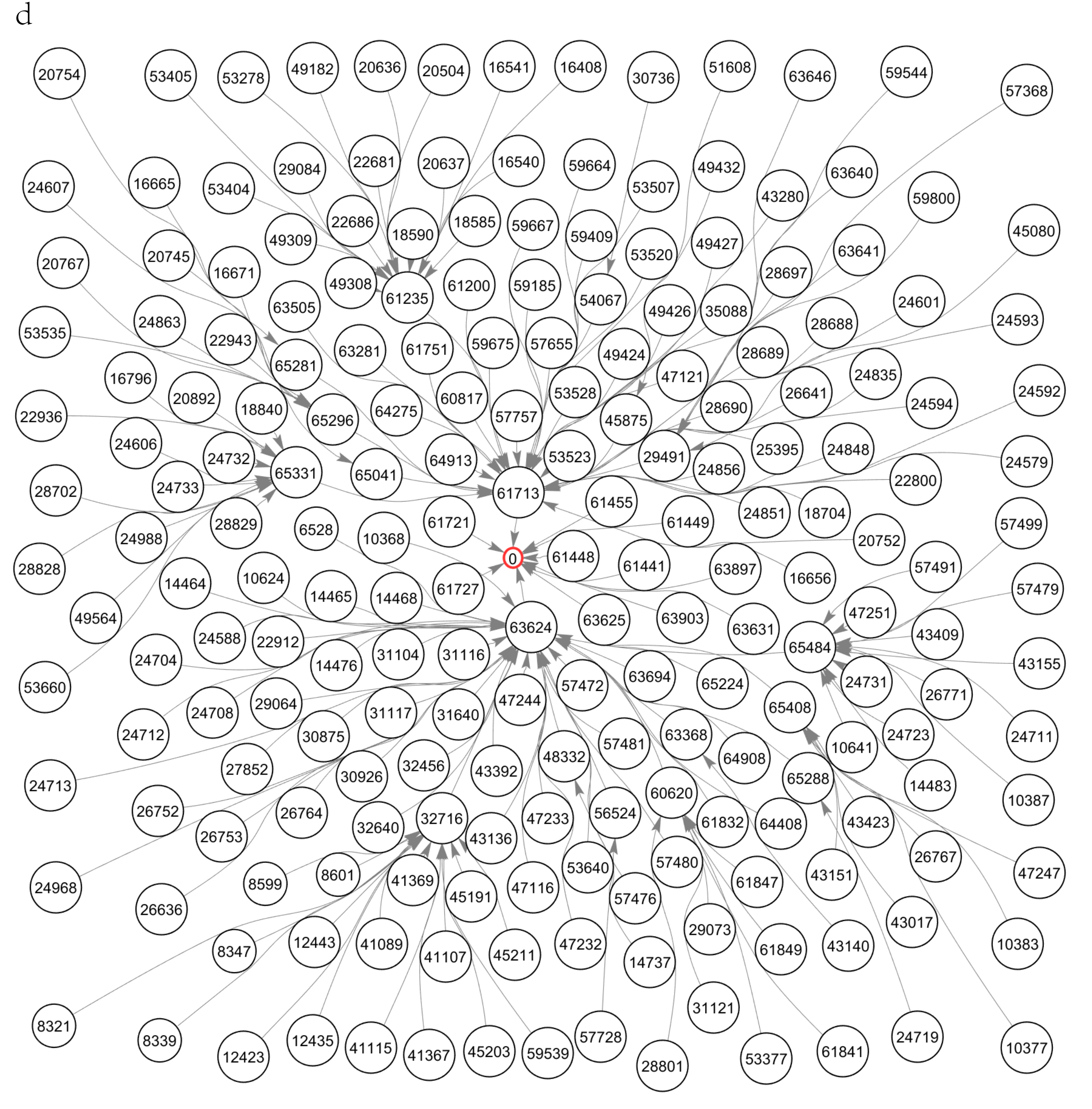}}
	%\caption
\begin{quote}
\textbf{Figure S2.} Class of states converging to full cooperation. The full cooperation is colored Red and associated states would be eventually stablized at this state. Panels a, b, c and d each show a fraction of evolutionary paths converges to the full cooperation. These evolutionary paths have no overlap except the attractor of full cooperation. Each population configuration is transformed into one state following the same way as in Figure 1.
\end{quote}
	\label{level}
\end{figure}

\newpage
\begin{figure}
  \centering
  \includegraphics[scale=2]{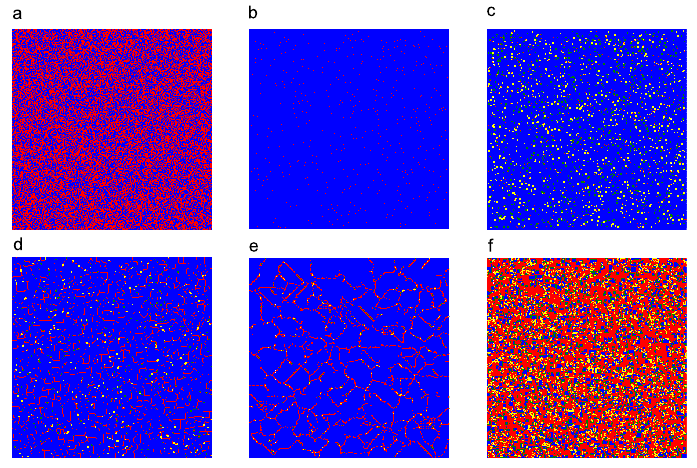}
  %\caption{
\begin{quote}
  \textbf{Figure S3.} Evolutionary dynamics for spatial Prisoner's Dilemma when the parameter of $b$ lies in different intervals. For $1<b<9/8$, a few defectors surviving the evolutionary race are totally in isolation. In stable state, no individuals switch strategy and so the pattern is static (b). For $9/8<b<8/7$, a few defectors each are surrounded by eight neighbors who switch between cooperation and defection periodically (c). For $8/7<b<1.8(b\notin \mathcal{ZP})$, defectors are able to form lines and separate defector clusters of size $3 \times 3$ reduce.  The larger $b$, the more tightly connected the lines, the fewer the separate defector clusters. Defectors cannot expand into larger clusters. The evolution can still quickly converge. Pattern d is for $7/5<b<3/2$ and pattern e is for $7/4<b<9/5$. For $1.8<b<2$, the periodical cycle is still not found in $1500000$ generations (see Fig. S3f). Pattern f is derived at generation $400000$. All five evolutionary processes start with the same initial configuration, in which $50\%$ cooperators are randomly distributed in a square-lattice population of size $400\times 400$ with fixed boundary conditions (pattern a).
\end{quote}
\end{figure}

\newpage
\begin{figure}
  \centering
  \includegraphics[scale=2]{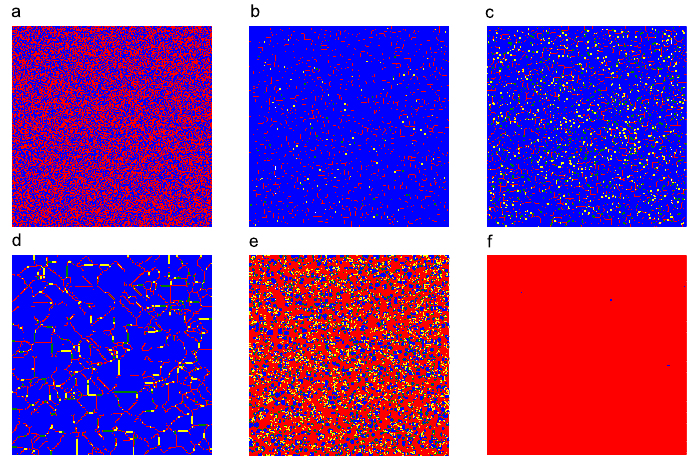}
%  \caption{
\begin{quote}
  \textbf{Figure S4.} When self-interaction is precluded, evolutionary processes can quickly enter periodical cycles (patterns b, c, d, f) except for $8/5<b<5/3$. For this interval, cooperators form clusters and defectors form clusters. Both kinds of clusters collide, transform, and rotate, and consequently a great number of patterns are visited. Though the pattern transits over generations deterministically through viewing the population as a whole, it is still greatly time-consuming to verify if these patterns visited are already inside periodical cycle or still in transient phase (pattern e). All five evolutionary processes start with the same initial configuration, in which $50\%$ cooperators are randomly distributed in a square-lattice population of size $400\times 400$ with fixed boundary conditions (pattern a).
\end{quote}
\end{figure}
\end{document}